\documentclass[twocolumn,preprintnumbers,amsmath,amssymb,superscriptaddress]{revtex4}
\usepackage{graphicx}
\usepackage{dcolumn}
\usepackage{bm}
\usepackage{soul}
\usepackage{color}
\usepackage{epstopdf}
\usepackage[version=3]{mhchem}
\usepackage{lipsum}
\usepackage[outercaption]{sidecap}
\usepackage{floatrow}
\usepackage{hyperref}

\begin{document}


\title{Effects of Mid-infrared Graphene Plasmons on Photothermal Heating}

\author{Anh D. Phan}
\email{anh.phanduc@phenikaa-uni.edu.vn}
\affiliation{Phenikaa Institute for Advanced Study, Artificial Intelligence Laboratory, Faculty of Information Technology, Materials Science and Engineering, Phenikaa University, Hanoi 12116, Vietnam}
\affiliation{Department of Nanotechnology for Sustainable Energy, School of Science and Technology, Kwansei Gakuin University, Sanda, Hyogo 669-1337, Japan}
\author{Do T. Nga}
\affiliation{Institute of Physics
Vietnam Academy of Science and Technology,
10 Dao Tan, Ba Dinh, Hanoi 12116, Vietnam}
\author{Do C. Nghia}
\affiliation{Hanoi Pedagogical University 2, Nguyen Van Linh Street, Vinh Phuc, Vietnam}
\author{Vu D. Lam}
\affiliation{Graduate University of Science and Technology, Vietnam Academy of Science and Technology, 18 Hoang Quoc Viet, Hanoi, Vietnam}
\author{Katsunori Wakabayashi}
\affiliation{Department of Nanotechnology for Sustainable Energy, School of Science and Technology, Kwansei Gakuin University, Sanda, Hyogo 669-1337, Japan}
\date{\today}

\begin{abstract}
We theoretically investigate the plasmonic heating of graphene-based systems under the mid-infrared laser irradiation, where periodic arrays of graphene plasmonic resonators are placed on dielectric thin films. Optical resonances are sensitive to structural parameters and the number of graphene layers. Under mid-infrared laser irradiation, the steady-state temperature gradients are calculated. We find that graphene plasmons significantly enhance the confinement of electromagnetic fields in the system and lead to a large temperature rise compared to the case without graphene. The correlations between temperature change and the optical power, laser spot, and thermal conductivity of dielectric layer in these systems are discussed. Our numerical results are in accordance with experiments.
\end{abstract}

\maketitle
Photothermal effects have been of interest in efficient solar vapor generation \cite{17}, optical data storage \cite{18}, alternative cancer therapy \cite{21,22}, antibacterial activities \cite{23}, and radiative cooling \cite{19}. The incident light excites surface plasmons of metal-like materials in photothermal agents and thus is effectively confined in the nanostructures. The confinement of electromagnetic waves causes strong light-absorbance. The absorbed energy converts efficiently into heat. Manipulating the electron density or plasmonic structures in systems leads to an inhomogeneity of electric fields and generates local heat to match with desired purposes. Typically, photothermal applications have employed noble metals because of their large concentration of free electrons and high efficiency of light-to-heat conversion \cite{24,25}. However, large inelastic losses of noble metals limit plasmon lifetimes in metal nanostructures and reduce service life of optical confinement.


Graphene has been recently considered as a novel plasmonic material \cite{2,3,4}, which strongly confines electromagnetic fields but provides relatively low loss \cite{20}. The optical and electrical properties of graphene can be easily tuned by doping, applying external fields, and injecting charge carriers \cite{4}. The small number of free electrons in graphene leads to occurance of plasmon resonance at the infrared regime and significantly reduces the heat dissipation compared to metals. Consequently, graphene-based metamaterials are of interest and potentially display various intriguing behaviors. 

There are several methods to investigate the performance of graphene-based devices. While experimental implementation is very expensive and simulation is time-consuming, theoretical approaches are scalable and provide good insights into the physics underlying design of metamaterials. Furthermore, the scalability is important in generating huge amount of data for machine learning and deep learning analyses. Thus, reliable theoretical models are useful for enhancing the highly accurate fabrications in both size and morphologies.

In this letters, we present a theoretical approach to calculate plasmonic properties of graphene-based nanostructures and temperature distribution in the systems when irradiated by a mid-infrared laser light. The systems are modeled to be similar to mid-infrared graphene detectors fabricated in Ref. \cite{2}. These calculations allow us to determine roles of graphene plasmons on the light confinement and heat generation. In addition, effects of dielectric layers on the temperature gradient are also discussed.

\begin{figure}[htp]
\includegraphics[width=8.2cm]{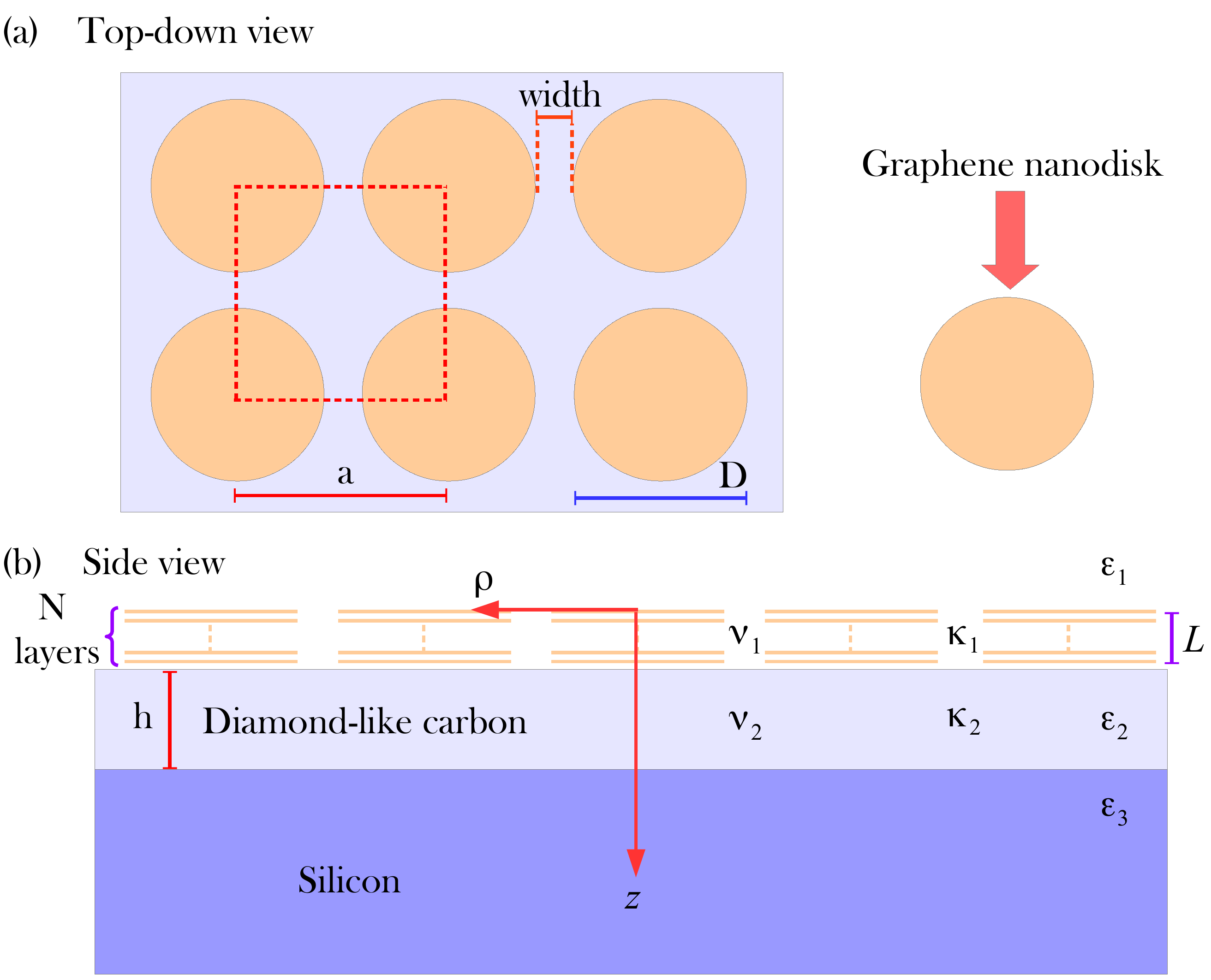}
\caption{\label{fig:1}(Color online) (a) The top-down view and (b) the side view of graphene based system including structural parameters.}
\end{figure}

Our graphene-based systems as depicted in Figure \ref{fig:1} composed of a square lattice of graphene nanodisks on a diamond-like carbon thin film grown on a silicon substrate. The square lattice is characterized by a lattice period, $a=270$ nm, a resonator size, $D=210$ nm, and a number of graphene resonator layers, $N$. A thickness of the diamond-like carbon layer is $h = 60$ nm. Since the energy of optical phonon of the diamond is about 165 meV \cite{14}, one can expect that this energy of diamond-like carbon thin film has the same order. The value is higher than the phonon energy of \ce{SiO_2} (55 meV) \cite{27} and other conventional substrates. Consequently, the surface of diamond-like carbon films has a lower trap density which leads to be non-polar and chemically inert surface.

When the size is much smaller than the wavelength of incident light, we can apply quasi-static approximation to analytically determine the polarizability of a single graphene resonator \cite{1,15,4,5}. To capture effects of interactions between graphene resonators on the polarizability, the dipole approximation is employed for simplicity. This approximation works well as $a/D \geq 1.5$ \cite{28,29}. We suppose that the accuracy is still acceptable for $a/D \approx 1.29$ in our system, which describes the practical graphene detector in Ref. \cite{2}. 

\begin{eqnarray}
\alpha(\omega)=\frac{\varepsilon_1+\varepsilon_2}{2}D^3\frac{\zeta^2}{-i\omega D\cfrac{\varepsilon_1+\varepsilon_2}{2\sigma(\omega)}-\cfrac{1}{\eta}},
\label{eq:1}
\end{eqnarray}
where $\varepsilon_1 = 1$, $\varepsilon_2 = 6.25$, $\varepsilon_3 = 11.56$, $\omega$ is frequency, and $\zeta$ and $\eta$ are geometric parameters calculated using the lowest-order dipole plasmon model. The analytical dependence of $\zeta$ and $\eta$ on the morphology and thickness can be found in Ref. \cite{1}.

According to the random-phase approximation with zero-parallel wave vector, the $N$-layers graphene conductivity in mid-infrared regime is \cite{4}
\begin{eqnarray}
\sigma(\omega)=\frac{Ne^2 i |E_F|}{\pi\hbar^2\left(\omega + i\tau^{-1} \right)},
\label{eq:2}
\end{eqnarray}
where $e$ is the electron charge, $\hbar$ is the reduced Planck constant, $\tau$ is the carrier relaxation time, and $E_F$ is a chemical potential. Typically, $\sigma(\omega)$ contains both interband and intraband transitions. However, in the mid-infrared regime, the intraband conductivity is completely dominant. The in-plane optical conductivity of graphene in Eq. (\ref{eq:2}) ignores contributions of interband transitions. The approximation is only valid in the low frequency regime \cite{4}. To effectively combine Eq. (\ref{eq:1}) and (\ref{eq:2}), it is necessary to require the strong coupling condition \cite{12,13}, where the spacing distance between graphene disks is much smaller than the diameter $D$. In addition, since graphene disk layers are separated by a dielectric layer \cite{13}, the vertical optical conductivity is ignored and the horizontal optical conductivity is simply additive.

The reflection and transmission coefficient of the graphene-based nanostructure are \cite{4,5}
\begin{eqnarray}
t_{13} &=& \frac{t_{12}t_{23}e^{i\left( \frac{\omega}{c}\sqrt{\varepsilon_2}h\right)}}{1+r_{12}r_{23}e^{2i\left( \frac{\omega}{c}\sqrt{\varepsilon_2}h\right)}},
\nonumber\\
r_{13} &=&  \frac{r_{12}+r_{23}e^{2i\left( \frac{\omega}{c}\sqrt{\varepsilon_2}h\right)}}{1+r_{12}r_{23}e^{2i\left( \frac{\omega}{c}\sqrt{\varepsilon_2}h\right)}}
\label{eq:3}
\end{eqnarray}
where
\begin{eqnarray}
r_0 &=& \frac{\sqrt{\varepsilon_2}-\sqrt{\varepsilon_1}}{\sqrt{\varepsilon_2}+\sqrt{\varepsilon_1}}, \quad t_0 = \frac{2\sqrt{\varepsilon_1}}{\sqrt{\varepsilon_2}+\sqrt{\varepsilon_1}}, \nonumber\\
r_{12} &=& r_0 - \frac{is(1-r_0)}{\alpha^{-1}-\gamma}, \quad t_{12} = t_0 + \frac{ist_0}{\alpha^{-1}-\gamma}, \nonumber\\
r_{23} &=& \frac{\sqrt{\varepsilon_3}-\sqrt{\varepsilon_{22}}}{\sqrt{\varepsilon_3}+\sqrt{\varepsilon_2}}, \quad t_{23} = \frac{2\sqrt{\varepsilon_2}}{\sqrt{\varepsilon_3}+\sqrt{\varepsilon_2}}, \nonumber\\
s &=& \frac{4\pi}{a^2}\frac{\omega/c}{\sqrt{\varepsilon_2}+\sqrt{\varepsilon_1}}, \quad \gamma \approx \frac{g}{a^3}\frac{2}{\varepsilon_1+\varepsilon_2}+is,
\label{eq:4}
\end{eqnarray}
where $c$ is the speed of light, $g\approx 4.52$ is the net dipolar interaction over the whole square lattice, $r_{pq}$ and $t_{pq}$ are the bulk reflection and transmission coefficients, respectively, when electromagnetic fields strike from medium $p$ to $q$. From these, we compute the transmission $|t_{13}|^2 \equiv |t_{13}(N)|^2$ for $N > 0$ and $N = 0$ corresponding to systems with and without graphene plasmonic resonators. The extinction spectra measured in the experiments is the relative difference in these transmissions 1-$|t_{13}(N)|^2/t_{13}(N=0)|^2$. The calculations clearly determine confinement effects of electromagnetic fields due to graphene plasmons. 

Under infrared laser irradiation, the diamon-like carbon layers and graphene-disk resonators absorb electromagnetic energy and are heated up. The temperature rise in cylinderical coordinate, $\Delta T \equiv \Delta T(\rho,z)$, obeys the heat diffusion equation. Since the graphene resonator is a two-dimensional material, the thermal conductivity is anisotropic. This diffusion equation in the layer including all stacked plasmonic disks is
\begin{eqnarray}
\kappa_{1\parallel}\frac{1}{\rho}\frac{d}{d\rho}\left(\rho \frac{d \Delta T}{d\rho}\right) + \kappa_{1\perp}\frac{d^2\Delta T}{dz^2}=p_0e^{-\frac{2\rho^2}{w^2}}e^{-\nu_1 z},
\label{eq:5}
\end{eqnarray}
where $p_0$ is the laser power per unit volume, $w$ is the laser spot, $\nu_1$ is the absorption coefficient, $\kappa_{1\parallel}$ and $\kappa_{1\perp}$ are the effective in-plane and out-of-plane thermal conductivity, respectively. To consider a uniform illumination condition, one simply takes $w=\infty$ and follows the same analysis as in Ref. \cite{31}.

\begin{figure}[htp]
\includegraphics[width=8.4cm]{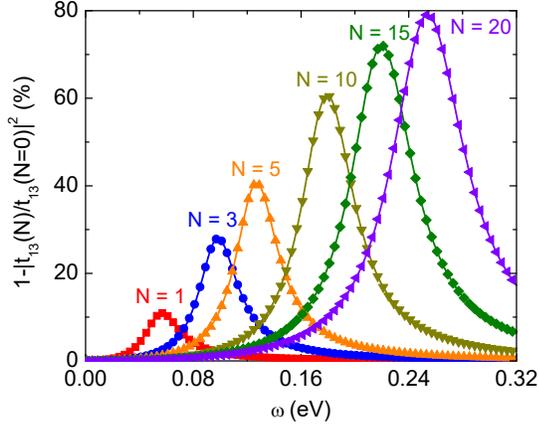}
\caption{\label{fig:2}(Color online) Theoretical  extinction spectra for systems having a graphene-disk array with $E_F = 0.45$ eV and $\hbar\tau^{-1}=0.03$ eV at several numbers of graphene layers.}
\end{figure}

These conductivities are strongly dependent on the fraction of graphene plasmons in the plasmonic layer, which is $f_p=\pi D^2/4a^2$. By adopting Maxwell-Garnett (MG) theory for an effective medium approximation, $\kappa_{1\parallel}$ and $\kappa_{1\perp}$ are 
\begin{eqnarray}
\kappa_{1\parallel,\perp} &=& \kappa_m\frac{(1-f_p)(\kappa_{\parallel,\perp}^{bulk}+2\kappa_m)+3f_p\kappa_{\parallel,\perp}^{bulk}}{(1-f_p)(\kappa_{\parallel,\perp}^{bulk}+2\kappa_m)+3f_p\kappa_{m}},
\label{eq:12}
\end{eqnarray}
where $\kappa_{\parallel}^{bulk} \approx 630$ $W/m/K$ \cite{6,8} and $\kappa_{\perp}^{bulk} \approx 6$ $W/m/K$ \cite{8} are the corresponding conductivity of bulk graphene, and $\kappa_{m} \approx 0.6$ $W/m/K$ is the thermal conductivity of medium (air). Here, we ignore effects of  plasmon and phonon coupling between graphene layers. A prior work \cite{30} indicated that the vertical coupling of graphene nanodisks is relatively strong and plays an important role in the frequency range from 140 $\ce{cm^{-1}}$ to 350 $\ce{cm^{-1}}$. However, resonant frequencies in our optical spectra in Fig. \ref{fig:2} are greater than 0.57 eV ($\sim 460$ $\ce{cm^{-1}}$). Thus, this strong coupling between vertically stacked graphene nanodisks may not (or may) happen in this work.

According to Beer-Lambert's law, one can estimate $\nu_1\approx Q_{abs}/a^2L$, in which $Q_{abs}=4\sqrt{2}\pi\omega Im(\alpha(\omega))/c\sqrt{\varepsilon_1+\varepsilon_2}$ is the absorption cross section, $L=N\delta$ is the thickness of graphene nanostructures with $\delta = 0.335$ nm, and $1/(a^2L)$ is the density of graphene resonators on the surface of the dielectric film. In the diamond-like layer, the diffusion equation is
\begin{eqnarray}
\kappa_{2}\left[\frac{1}{\rho}\frac{d}{d\rho}\left(\rho \frac{d \Delta T}{d\rho}\right) + \frac{d^2\Delta T}{dz^2}\right]= p_0e^{-\frac{2\rho^2}{w^2}}e^{-\nu_1L-\nu_2(z-L)},
\label{eq:6}
\end{eqnarray}
where $\kappa_2 \approx 0.6$ $W/K/m$ \cite{7} is the thermal conductivity of the diamond-like layer and $\nu_2 \approx 1.5$ $\mu m^{-1}$ is the absorption coefficient. For simplification purpose, we assume that the silicon substrate is kept at ambient temperature by contacting with a thermostat. This boundary condition was used to successfully analyze experiments in Ref. \cite{16}.

To solve these differential equations, we take the Hankel transform of the above equations in $\rho$ and it gives
\begin{widetext}
\begin{eqnarray}
-\kappa_{1\parallel}u^2\Theta(u,z) + \kappa_{1\perp}\frac{d^2\Theta(u,z)}{dz^2}&=&\frac{\nu_1P_0(1-R)}{2\pi}e^{-\frac{u^2w^2}{8}}e^{-\nu_1 z}, \qquad 0\leq z \leq L\\
\label{eq:7}
\kappa_{2}\left[-u^2\Theta(u,z) + \frac{d^2\Theta(u,z)}{dz^2}\right]&=&\frac{\nu_2P_0(1-R)}{2\pi}e^{-\frac{u^2w^2}{8}}e^{-\nu_1 L}e^{-\nu_2(z-L)}, \qquad L\leq z \leq L+  h
\label{eq:8}
\end{eqnarray}
\end{widetext}
where $\Delta T(\rho,z)=\int_0^\infty\Theta(u,z)J_0(\rho u)udu$ with $J_0$ being the Bessel function of the first kind. $P_0$ is the power of the incident flux. A correction factor $(1-R)$ implies that only the absorption and transmission component of light play a role in the heating process. The analysis is consistent with a recent work \cite{31}. The reflection is calculated by $R = \left|r_{13}\right|^2$. At $\omega \approx 0.1$ eV, except for $R \approx 0.376$ as $N=3$, $R \approx 0.3$ when $N$ changes from 0 to 10.

By solving Eqs. (8) and (\ref{eq:8}), one obtains the solutions
\begin{widetext}
\begin{eqnarray}
\Theta(u,z) &=& A_1(u)e^{-\sqrt{\frac{\kappa_{1\parallel}}{\kappa_{1\perp}}}uz}+B_1(u)e^{\sqrt{\frac{\kappa_{1\parallel}}{\kappa_{1\perp}}}uz}+ \frac{\nu_1P_0(1-R)}{2\pi(k_{1\parallel}u^2-\kappa_{1\perp}\nu_1^2)}e^{-\frac{u^2w^2}{8}}e^{-\nu_1 z}, \qquad 0\leq z \leq L\\
\label{eq:9}
\Theta(u,z) &=& A_2(u)e^{-uz}+B_2(u)e^{uz}+ \frac{\nu_2P_0(1-R)}{2\pi\kappa_2(u^2-\nu_2^2)}e^{-\frac{u^2w^2}{8}}e^{-\nu_1L}e^{-\nu_2(z-L)}, \qquad L\leq z \leq L+  h
\label{eq:10}
\end{eqnarray}
\end{widetext}
where $A_1(u)$, $B_1(u)$, $A_2(u)$, and $B_2(u)$ are parameters determined by boundary conditions
\begin{eqnarray}
-\kappa_{1\perp}\left.\frac{\partial \Theta(u,z)}{\partial z}\right|_{z=0} &=& 0, \nonumber\\
\Theta(u,L^-) &=& \Theta(u,L^+), \nonumber\\
-\kappa_{1\perp}\left.\frac{\partial \Theta(u,z)}{\partial z}\right|_{z=L^-} &=& -\kappa_{2}\left.\frac{\partial \Theta(u,z)}{\partial z}\right|_{z=L^+}, \nonumber\\
\Theta(u,z=L+h) &=& 0.
\label{eq:11}
\end{eqnarray}

Figure \ref{fig:2} shows theoretical infrared extinction spectra of graphene-based systems with several values of graphene plasmon layers calculated using Eqs. (\ref{eq:3}) and (\ref{eq:4}). The numerical results indicate the plasmonic peak for a square lattice of three-layers-graphene disks is roughly located at 0.1 eV ($\sim 806$ $\ce{cm^{-1}}$), which quantitatively agrees with experiment in Ref. \cite{2}. The presence of graphene plasmons reduces the transmission of electromagnetic fields through these systems. An inrease of $N$ blue-shifts the plasmonic resonance and enhances the amplitude signal in the optical spectra. More mid-infrared optical energy is confined in the system as increasing the layer number of graphene plasmons. The amount of the trapped energy can be indirectly measured via temperature caused by the light-to-heat conversion process. 

Other parameters can tuned to modify the extinction spectrum. Since $\alpha(\omega) \sim D^3$, a reduction of the diameter weakens the in-plane plasmonic coupling among resonators and lowers the optical peak when fixing $a$. Similar behaviors can be observed if the diameter remains unchanged and we increase the lattice period $a$. Meanwhile due to $\sigma(\omega) \sim E_F$, effects of graphene plasmons on the optical spectrum becomes less important when decreasing $E_F$. 

\begin{figure}[htp]
\includegraphics[width=8.4cm]{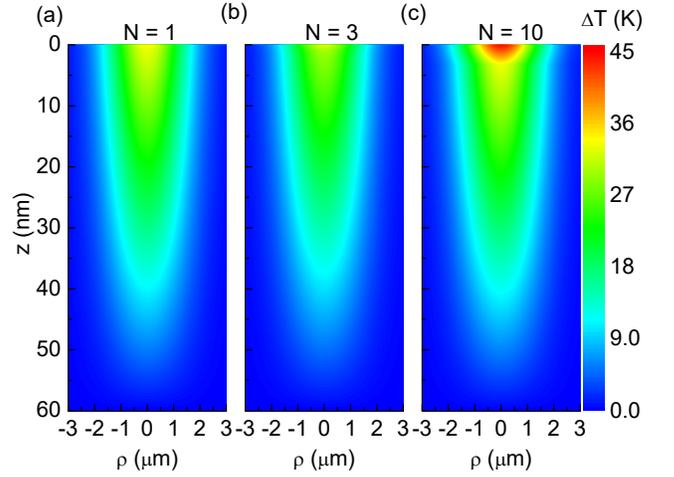}
\caption{\label{fig:5}(Color online) Spatial contour plots of the steady-state temperature rise in Kelvin units in graphene-based systems having (a) $N = 1$, (b) $N=3$, and (c) $N=10$ under illumination of a quantum cascade laser light.}
\end{figure}

To investigate effects of the graphene plasmon layers on the heating of graphene-disk systems, we use Eq. (10), (\ref{eq:10}) and (\ref{eq:11}) to calculate the spatial distributions of steady-state temperature when exposed by an infrared laser beam. This laser operates at $\sim 0.1$ eV with the incoming power $P_0=630$ $\mu W$ and $w=2.3$ $\mu m$. Numerical results are shown in Fig. \ref{fig:5} for different $N$. The incident photons are highly localized at the surface and spatially decay towards the bottom of the diamond-like carbon layer. The temperature increase at the hottest spot area ($z = 0$ and $\rho = 0$) in the case of $N = 10$ is 45 $K$, while that of systems having $N=3$ and $N=1$ are approximately 33 $K$ and 34.5 $K$, respectively. These temperature rises are much higher than $\Delta T(\rho=0,z=0)\approx 1.67$ $K$ for the system without graphene plasmons. This result clearly indicates that the dielectric loss is dominated by the ohmic loss (Joule heating) on graphene resonators under the mid-infrared irradiation.

The temperature rise strongly depends on the incident power $P_0$ and the laser spot $w$. On the basis of Eqs. (10), (\ref{eq:10}) and (\ref{eq:11}), one analytically finds that $\Delta T(\rho,z)$ is proportional to $P_0$. This finding is consistent with many prior works \cite{22,6,26}. For $N=3$, our numerical results give $\Delta T(0,0)=0.0521 P_0$ and predicts $\Delta T(0,0)\approx 330$ $K$ as $P_0 = 6.3$ $mW$, which is close to $\Delta T \approx 365$ $K$ \cite{6}. Meanwhile, at fixed incoming power, an increase of $w$ leads to a decrease of the laser intensity ($2P_0/\pi w^2$). It is well-known that the absorbed optical energy of a plasmonic resonator is approximately $2Q_{abs}P_0/\pi w^2$. This analysis suggests that  $\Delta T_{max} \sim 1/w^2$ and/or $\log_{10}(\Delta T(0,0))$ is inversely linear to $\log_{10}(w)$. This correlation is in a good agreement with numerical calculations.

\begin{figure}[htp]
\includegraphics[width=8.4cm]{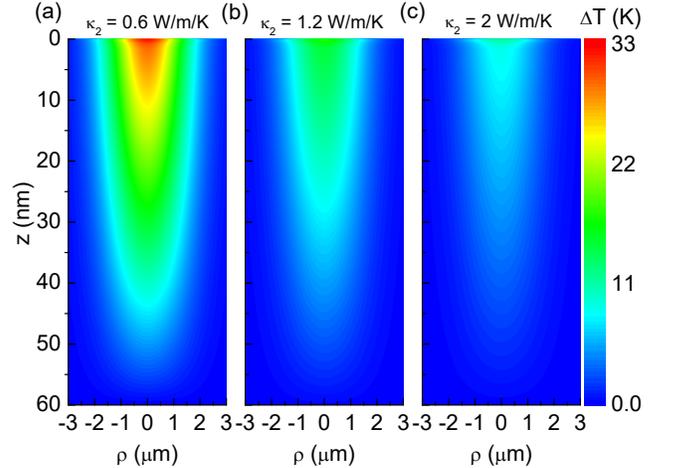}
\caption{\label{fig:6}(Color online) Spatial contour plots of the steady-state temperature rise in Kelvin units in graphene-based systems having $N = 3$ and (a) $\kappa_2=0.6$ $W/m/K$, (b) $\kappa_2=1.2$ $W/m/K$, and (c) $\kappa_2=2.0$ $W/m/K$ under illumination of a quantum cascade laser light.}
\end{figure}

The thermal conductivity of the thin film ($\kappa_2$) below graphene plasmons has a significant influence on the temperature rise of graphene-based systems. To zeroth-order approximation, we assume a decoupling between $\kappa_2$ and $\varepsilon_2$. This assumption may be reasonable since the phonon scattering, which is very important in the lattice thermal conductivity, strongly depends on the interfacial roughness \cite{9,10}. Effects of the grain boundary on the thermal conductivity is frequency independent \cite{9,10}. While a prior work \cite{11} indicated that the grain boundary does not alter the dielectric and infrared response of a ceramic material at room temperature. Thus, one can modify the grain boundary scattering without changing the dielectric constant of the host material.

Figure \ref{fig:6} shows that the steady temperature profile of the system for different values of $\kappa_2$ illuminated by the mid-infrared laser light. We use the same laser as the calculations in Figure \ref{fig:5}. The object having a larger $\kappa_2$ requires more thermal energy to be heated. Thus, the temperature rise is depressed.

We have proposed to investigate the plasmonic heating of graphene-based systems under irradiation of a mid-infrared laser. The nanostructures comprise a square array of multilayer graphene nanodisks deposited on the diamond-like carbon thin film, which is supported by a silicon substrate. We employ the dipole model associated with the random-phase approximation to calculate the polarizability of graphene on the substance and its absorption cross section. From these, the reflection and transmission coefficients are theoretically computed to estimate the confined power of the incident light. Our numerical results are in accordance with experiments \cite{2}. When illuminating the systems by the laser light, plasmonic nanodisks absorb more optical energy than the counterparts without graphene and convert to thermal dissipation. This finding indicates that the Ohmic loss is much larger than the dielectric loss in the mid-infrared regime. An increase of graphene plasmonic layers enhances the thermal gradients. At fixed number of graphene layers, the temperature rise is linearly proportional to the optical power and decays as the inverse square of the laser spot. Furthermore, a decrease of the heated temperature as increasing the thermal conductivity of the thin film layer is also calcualted and discussed.
\begin{acknowledgments}
This research is funded by Vietnam National Foundation for Science and Technology Development (NAFOSTED) under grant number 103.01-2018.337. This work was supported by JSPS KAKENHI Grant Numbers JP19F18322 and JP18H01154.

Conflict of Interest: The authors declare that they have no conflict of interest.
\end{acknowledgments}

\end{document}